\documentclass[fleqn,twoside]{article}
\usepackage{espcrc2}

\usepackage{graphicx}

\newcommand{\AmS}{{\protect\the\textfont2
  A\kern-.1667em\lower.5ex\hbox{M}\kern-.125emS}}

\title{Vacuum stress around a topological defect}

\author{V. A. De Lorenci\address{Instituto de Ci\^encias,
Universidade Federal de Itajub\'a,\\
Av.\ BPS 1303 Pinheirinho, 37500-903 Itajub\'a, MG, Brazil}%
        \thanks{delorenci@unifei.edu.br}
        and
        E. S. Moreira, Jr.\addressmark\thanks{moreira@unifei.edu.br}}
       
\begin{document}

\begin{abstract}
We show that a dispiration (a disclination plus a screw dislocation)
polarizes the vacuum of a scalar field giving rise to an energy
momentum tensor which, as seen from a local inertial  frame,
presents non vanishing off-diagonal components.
The results may have applications in cosmology 
(chiral cosmic strings) and condensed matter physics (materials with
linear defects). 
\vspace{1pc}
\end{abstract}

\maketitle

%
%
It is fairly well known that a needle solenoid carrying
a magnetic flux makes virtual charged particles to run 
around the solenoid inducing a
non vanishing current density (see e.g. Ref \cite{ser85}).  
We wish to consider what seems to be  a gravitational
(geometric) analogue of  this Aharonov-Bohm effect,
by computing the vacuum expectation value of the energy momentum
tensor of a massless and neutral scalar field far away from a dispiration.

%
%
Let us begin by presenting the geometry of the background
(units are such that $c=\hbar=1$),
\begin{equation}
ds^2=dt^2 - dr^2 - \alpha^2 r^2 d\theta^2 - (dz + \kappa d\theta)^2,
\label{2}
\end{equation}
where the points labeled by 
$(t,r,\theta,z)$ and $(t,r,\theta +2\pi,z)$ are identified 
\cite{gal93,tod94}.
When $\alpha=1$ and $\kappa=0$ 
Eq (\ref{2}) becomes the
line element of the flat spacetime written in cylindrical coordinates.
Borrowing terminologies in condensed matter physics, the parameters
$\alpha$ and $\kappa$ correspond to a disclination and a screw dislocation,
respectively. 
We should remark  that Eq (\ref{2}) may be  associated with   the gravitational
background of certain chiral cosmic strings \cite{bek92}
(as has been suggested in Ref. \cite{gal93}), 
as well as can describe (in the continuum limit)  the effective geometry
around a dispiration in an elastic solid (see Ref. \cite{pun97} and references
therein).

The definitions
$\varphi:= \alpha\theta$ and $Z := z + \kappa\theta$ lead to  
\begin{equation}
ds^{2}=dt^{2}-dr^2 -r^2 d\varphi^2 - dZ^2,
\label{mmet}
\end{equation}
which should be considered together with the  peculiar identification
\begin{equation}
(t,r,\varphi,Z) \sim (t,r,\varphi+2\pi\alpha,Z+2\pi\kappa).
\label{id}
\end{equation}
Although Eq. (\ref{mmet}) expresses the fact that the background is 
locally flat, due to Eq. (\ref{id}) we cannot use Eq. (\ref{mmet})
(which is a local statement) to infer that 
the global symmetries of the background are the same as those 
of  the Minkowski spacetime (in this sense Eq. (\ref{mmet}) is singular).
In fact, Eq. (\ref{mmet}) disguises  a curvature singularity on the symmetry
axis \cite{gal93}
(when $\kappa\neq 0$, in the context of the Einstein-Cartan theory, 
there is also a torsion singularity at $r=0$
\cite{tod94,let95}).

%
%
%

The vacuum expectation value of the energy momentum tensor is
obtained by applying a differential operator 
to the renormalized scalar propagator around a dispiration 
(see e.g. Ref. \cite{dav82}),
\begin{equation}
\left<T^\mu{}_\nu\right> = i \lim_{x'\rightarrow x}
{\cal D}^\mu{}_\nu(x,x')\
D^{(\alpha,\kappa)}(x,x').
\label{prescription}
\end{equation}
We have recently obtained
$D^{(\alpha,\kappa)}(x,x')$
(classical propagators have been considered in Ref \cite{lor02}) by
using the Schwinger proper time prescription combined with the 
completeness relation of the eigenfunctions of the d'Alembertian operator
\cite{lor03}. Such eigenfunctions have the form
$R(r)\chi(\varphi)\exp\{i(\nu Z-\omega t)\}$ 
which, by observing Eq. (\ref{id}), leads to 
\begin{equation}
\chi(\varphi +2\pi\alpha)= e^{-i2\pi\nu\kappa}\chi(\varphi). 
\label{abbcondition}
\end{equation}
This boundary condition is typical of the Aharonov-Bohm set up
where $\nu\kappa$ is identified with the flux parameter $e\Phi/2\pi$.
If we carry over to the four-dimensional context lessons from 
gravity in three dimensions \cite{ger89a,ger90}, it follows that
the charge $e$ and the magnetic flux $\Phi$ should be identified with
the longitudinal linear momentum $\nu$ and $2\pi\kappa$, respectively
\cite{gal93}.  

When $\kappa/r\rightarrow 0$,  Eq. (\ref{prescription}) yields
for the diagonal components the expressions 
of the vacuum fluctuations around 
an ordinary cosmic string ($\kappa=0$) \cite{emi94}.
Regarding the other components, the prescription in Eq.
(\ref{prescription}) kills off the dominant contribution in the renormalized
propagator \cite{lor03}, resulting that the subleading contribution yields two non
vanishing off-diagonal components,  
\begin{equation}
\left<T^{\varphi}{}_{Z}\right> = \frac{i}{r^2}\lim_{x'\rightarrow x}
\partial_{\varphi}\partial_{Z}D^{(\alpha,\kappa)}(x,x')
= \frac{\kappa}{r^6} B(\alpha),
\label{t23}
\end{equation}
and 
\begin{equation}
\left<T^{Z}{}_{\varphi}\right> 
= \frac{\kappa}{r^4} B(\alpha),
\label{t32}
\end{equation}
where $B(\alpha)$ depends on the disclination parameter only \cite{lor03}. 
Unlike the diagonal components,
$\left<T^{\varphi}{}_{Z}\right>$
and  
$\left<T^{Z}{}_{\varphi}\right>$ 
do not depend on the coupling parameter $\xi$.  

When $\alpha=1$, $B=1/60\pi^{2}$ which
corresponds approximately to the value of $\alpha$ in the physics
of formation of ordinary cosmic strings \cite{vil94}.   

It is instructive to display both disclination and screw dislocation
effects in a same array. When $\xi=1/6$ (conformal coupling), for 
example, $\left< T^{\mu}{}_{\nu} \right>$  with respect to the 
local inertial frame [cf. Eq. (\ref{mmet})] can
be cast into the form  
\begin{equation}
\left< T^{\mu}{}_{\nu} \right> =\frac{1}{r^4}
\left(
     \begin{array}{cccc}
      -A &  0 & 0                    & 0 \\
       0 & -A & 0                    & 0 \\
       0 &  0 & 3A                   & \kappa B/r^2 \\
       0 &  0 & \kappa B             & -A
     \end{array}
\right)
\label{tmunumatrix},
\end{equation}
where $A(\alpha):=(\alpha^{-4}-1)/1440\pi^2$, and which holds 
far away from the defect 
(and for $\alpha\neq 1$, when $\kappa\neq 0$).
[When $\kappa\neq 0$, by setting $\alpha=1$ in Eq. (\ref{tmunumatrix}), 
$A$ vanishes and subleading contributions depending on $\kappa$ take over.] 

Before closing we should report a discrepancy in the literature. Recently,
works \cite{pont98,hell03} have appeared stating that, for $\alpha=1$,
the subleading contribution in $\left< T^0{}_0\right>$ 
[cf. Eq. (\ref{tmunumatrix})] is $\lambda \kappa^2 / r^6$, where 
$\lambda$ is a certain constant. We have performed a rather detailed
study showing that such a subleading contribution is 
$f(\kappa/r)\,\kappa^2/r^6$ instead, where $f(\kappa/r)$ diverges as 
$\kappa\rightarrow 0$ [although $\lim_{\kappa\rightarrow 0}
\kappa^2 f(\kappa/r)=0$].

\section*{acknowledgments}
This work was partially supported by the Brazilian research agencies
CNPq and FAPEMIG.

\end{document}